\documentclass[aps,10pt,prd,twocolumn,eqsecnum,amssymb,amsmath,showpacs,a4paper,superscriptaddress,nofootinbib,floatfix]{revtex4-2}
\usepackage{import}

\usepackage{graphicx}
\usepackage{amsfonts}
\usepackage{amsmath}
\usepackage{hyperref}
\usepackage{units}
\usepackage{color}
\usepackage{dcolumn}% Align table columns on decimal point
\usepackage{bm} 
\usepackage{cleveref}
\usepackage{braket}
\usepackage{todonotes}
\crefname{section}{§}{§§}
\Crefname{section}{§}{§§}
\usepackage{mathtools}
\usepackage{chngcntr}
\counterwithout{equation}{section}
\usepackage[normalem]{ulem}
\usepackage{tikz}
\usepackage{comment}
\usepackage{nicefrac}
\usepackage{ragged2e}

\usepackage[T1]{fontenc}
\usepackage[utf8,applemac]{inputenc}
\newcommand{\be}{\begin{equation}}
\newcommand{\ee}{\end{equation}}
\newcommand{\bsub}{\begin{subequations}}
\newcommand{\esub}{\end{subequations}}
\newcommand{\bea}{\begin{eqnarray}}
\newcommand{\eea}{\end{eqnarray}}
\newcommand{\bi} {\begin{itemize}}
\newcommand{\ei} {\end{itemize}}
\newcommand{\bmat} {\begin{pmatrix}}
\newcommand{\emat} {\end{pmatrix}} 

\usepackage[final]{showlabels} %[inner] at end, [inline] in middle, [final] removes

\newcommand{\D}{\mathrm{d}}
\newcommand{\I}{\mathrm{i}}
\newcommand{\E}{\mathrm{e}}

\newcommand{\mrm}[1]{\mathrm{#1}}

\DeclareMathOperator{\sgn}{sgn}

\makeatletter
\newcommand*{\balancecolsandclearpage}{%
  \close@column@grid
  \clearpage
  \twocolumngrid
}
\makeatother

\makeatletter
\let\cat@comma@active\@empty
\makeatother

\usepackage{xcolor}
\hypersetup{  colorlinks,
              linktoc         = page, % section, page, all, none
              urlcolor        = {green!80!blue},
              linkcolor       = {orange!60!black}, % footnotes+ref's
              urlcolor        = {cyan!80!black}, % reference links
              citecolor       = {cyan!80!black}, % \cite's
              anchorcolor     = {yellow}
}

\begin{document}

\title{Testing Superpositions of Detector Trajectories}

% \title{Verifying Superpositions of Particle Detectors, or: Sensing Superpositions of Trajectories}

% Sensing Superpositions of Trajectories

% Sensing with Superpositions of Trajectories

\author{Cisco Gooding}
\affiliation{Laboratoire Kastler Brossel, Sorbonne Universit\'{e}, CNRS, ENS-Universit\'{e} PSL,  Coll\`{e}ge de France, 4 place Jussieu, F-75252 Paris, France}
\author{Taylor Cey}
\affiliation{Department of Physics and Astronomy, University of Waterloo,
200 University Ave W, Waterloo, Ontario N2L 3G1, Canada}
\author{Robert Mann}
\affiliation{Department of Physics and Astronomy, University of Waterloo,
200 University Ave W, Waterloo, Ontario N2L 3G1, Canada}
\affiliation{Perimeter Institute for Theoretical Physics, 31 Caroline St. N., Waterloo, Ontario N2L 2Y5, Canada}
\affiliation{Department of Applied Mathematics, University of Waterloo,
200 University Ave W, Waterloo, Ontario N2L 3G1, Canada}
\affiliation{
Institute for Quantum Computing, University of Waterloo,
200 University Ave W, Waterloo, Ontario N2L 3G1, Canada}
\date{\today}

\begin{abstract}
We propose a realizable experiment to test the response of a particle detector prepared in a superposition of locations interacting with a relativistic quantum field. Using a beamsplitter to prepare two superposed branches of a modulated laser probe, these branches are directed to intersect a pancake-shaped Bose-Einstein condensate at two separate locations. The branches are then recombined with another beamsplitter. Heterodyning one of the outputs, the response function corresponding to an Unruh-deWitt detector in a superposition of locations interacting with a $(2+1)$-dimensional massless scalar field is shown to appear in the difference photocurrent power spectrum. Operating beyond the standard quantum limit using squeezed light, the signal-to-noise ratio is estimated as $SNR\gtrsim 10$ for extracting the response function over a broad set of baseband frequencies.
\end{abstract}

\maketitle

\section{Introduction}

The study of quantum fields on both flat and curved spacetimes has been greatly facilitated by the use of particle-detector models, such as the ubiquitous Unruh-deWitt (UdW) model \cite{Unruh1979evaporation,DeWitt1979}. The pioneering work \cite{Unruh1979evaporation}, along with much subsequent work, has been devoted to a localized particle detector interacting with a quantum field on a fixed spacetime along a fixed classical trajectory. 

There has been growing interest in expanding the scope of the UdW model -- to include, for instance, superpositions of classical detector trajectories \cite{Foo:2020xqn}. 
The door is then open to study issues of causality and thermality with superpositions of accelerated trajectories
\cite{Foo:2020dzt}, as well as further questions about entanglement in flat and curved spacetimes \cite{Foo:2021gkl}. 
Expanding further this formalism, combined with additional methods from quantum reference frames \cite{Giacomini2019}, has natural utility as an operational approach to quantum gravity, to study superpositions of spacetimes \cite{Foo:2020jmi,Foo:2021exb,Foo:2022dnz,Suryaatmadja:2023onb}.

A recent experimental proposal aims to extract a signature of the response function for a particle detector governed by the UdW model, in the context of the circular motion Unruh effect, to be implemented in an analogue ultracold atom system \cite{Analog2,Analog1,Analog4}. Along with representing an experimental realization of the circular Unruh effect
\cite{Unruh1998}, this analogy affords the possibility of implementing an entanglement harvesting protocol \cite{Analog3}. 

 Here we make further use of the relativistic analogy to provide previously unavailable experimental access to the study of superpositions of detector trajectories. We propose an experimental 
test of the response of a particle detector in a superposition of locations interacting with a massless scalar field in $(2+1)$ dimensions. After deriving the predicted detector response function, we describe an experimental setup - shown schematically in Figure  \ref{fig:exp} - to realize the predicted response using a modulated laser and a pancake-shaped Bose-Einstein condensate (BEC). A beamsplitter is used to probe the BEC at two distinct interaction points, simulating a superposition of detector locations. We analyze the expected output from a heterodyne detection scheme, designed to extract a signal of the corresponding response function. The standard quantum limit for extracting this signal is identified, and the signal-to-noise ratio for optimally extracting the response function signal is estimated using a squeezed state for the initial laser probe. Finally, we discuss feasibility of experimental implementation, interpretation of the proposal in terms of quantum sensing, and the formulation from our results of a witness for detector superpositions.

\section{Unruh-deWitt Detectors}
\label{sec:UdW}

We start by considering a two-state particle detector that has a localized interaction with a massless scalar field $\hat{\phi}$ in $(2+1)$-dimensional Minkowski spacetime. The particle detector has energy eigenstates $\ket{g}$ and $\ket{e}$ with eigenvalues $0$ and $\nu$, respectively. The detector location is described by a quantum control state $\ket{\psi}$, which pertains to the trajectory the detector takes through spacetime, such that a coherent superposition of two trajectories can be expressed as 
\begin{equation}
    \ket{\psi} = \frac{1}{\sqrt{2}}\left(\ket{1}+\ket{2}\right)\, .
\end{equation}
The field $\hat{\phi}$ is prepared in the Minkowski vacuum state, and the particle detector is initialized in state $\ket{g}$. In the interaction picture, the Hamiltonian describing the interaction of the field with a UdW detector is
\begin{subequations}\label{eqn::Hints}
\begin{align}
     \hat{H}_{\mrm{int}} ( \tau)~=~& \hat{\mathcal{H}}_1(\tau)\otimes\ket{1}\bra{1}+ \hat{\mathcal{H}}_2(\tau)\otimes\ket{2}\bra{2}\,,\\
\hat{\mathcal{H}}_j(\tau)~=~&\lambda\,\hat \mu(\tau)\, \eta(\tau)\,\hat \phi(x_j(\tau))\,,
\end{align}    
\end{subequations}
where $\tau$ is the proper time along each trajectory $x_j(\tau)$ (for $j\in \{1,2\}$), $\lambda$ is the coupling constant, and $\eta(\tau)$ is the switching function. Here, $\hat \mu(\tau) $ is the monopole moment operator, given by
\begin{align}
    \hat \mu(\tau) 
    &=
        \ket{e} \bra{g} e^{ \I \nu \tau }
        +
        \ket{g} \bra{e} e^{ -\I \nu \tau }\, ,
\end{align}
with $\nu$ being the energy gap of the detector. For $\nu>0$ the state $\ket{e}$ is excited relative to the initial state $\ket{g}$, whereas for $\nu<0$ these roles are reversed. 

The transition probability for the detector is given by
\begin{equation}
    P(\nu) = \frac{\lambda^2}{4}\sum_{i=1}^2 \sum_{j=1}^2 P_{ij}(\nu),
    \label{eq:Psum}
\end{equation}
where
\begin{align}\label{eq:TransitionP_ij}
    P_{ij}(\nu) = \int d\tau_i\int d\tau_j'\: \eta(\tau_i)\eta(\tau_j') e^{-i\nu(\tau_i-\tau_j')} W_{ij}(\tau_i,\tau_j')
\end{align}
and
\begin{equation}
    W_{ij}(\tau_i, \tau_j') = \bra{0} \hat\phi(\mathsf{x}_i(\tau_i)) \hat \phi(\mathsf{x}_j(\tau_j')) \ket{0}
\end{equation}
is the vacuum Wightman function pulled back along the trajectories of the detector \cite{Foo:2020xqn}. For a massless scalar quantum field in $(2+1)$ dimensions pulled back to a single arbitrary trajectory $\mathsf{x}(s)$, the Wightnman function is given by \cite{UnruhTemps}
\begin{align}
    W(s,0)=\frac{1}{4\pi}\frac{1}{\sqrt{\left(\Delta \mathsf{x}(s-\I\epsilon)\right)^2}}\, ,
\end{align}
with the $\I\epsilon$ understood in the usual distributional sense. The Wightman function corresponding to a static detector is stationary; in this case the ``diagonal'' Wightman function reduces to
\begin{align}
        W_{jj}(s,0)\equiv W_{jj}(s)=\frac{1}{4\pi}\frac{1}{\sqrt{-(s-\I\epsilon)^2}}\, ,
\end{align}
where $s=\tau_j-\tau_j'$.

\section{Static Detector Superpositions}

In contrast, the off-diagonal Wightman function for a static detector in superposition of positions displaced by $\delta=|\bm{x}_1-\bm{x}_2|$ is 
\begin{align}\label{superWight}
    W_{ij}(s)=\frac{1}{4\pi}\frac{1}{\sqrt{-(s-\I\epsilon)^2+\delta^2}}\, ,
\end{align}
where $s=\tau_i-\tau_j'$ and $i\neq j$. Corresponding to the Wightman function \eqref{superWight}, using a constant switching function, is a response function
\begin{align}\label{offdiagF}
    \mathcal{F}_{ij}(\nu)=\int_{-\infty}^\infty \D s\, e^{-\I\nu s}\,W_{ij}(s)\, .
\end{align}
When the Wightman function is stationary, the response function for a detector is proportional to its transition rate \cite{UnruhTemps}. Care must be taken with the integration in determining the response function \eqref{offdiagF}, due to the branch points at $s=\I\epsilon\pm\delta$. Performing the contour integration, one finds
\begin{align}\label{offdiagrespo}
    \mathcal{F}_{ij}(\nu)=\frac{1}{2\pi}\Theta(-\nu)\int_{-\delta}^{\delta}\D z\, \frac{e^{-\I\nu z}}{\sqrt{\delta^2-z^2}}= \frac{1}{2}\Theta(-\nu)J_0(\nu\delta)\, ,
\end{align}
which reduces to the well-known single-location result
\begin{align}\label{singlimit}
    \mathcal{F}_{jj}(\nu)=\frac{1}{2}\Theta(-\nu)
\end{align}
in the $\delta\rightarrow 0$ limit \cite{UnruhTemps,Hodgkinson2012}.

\section{Relativistic Analogy}

We now describe a mathematical analogy between a massless Klein-Gordon field in $(2+1)$ dimensions and long-wavelength density fluctuations in a pancake-shaped BEC \cite{Analog1,PhysRevLett.85.4643,Analog2,Analog4}. The BEC has an effectively planar structure, obeying the $(2+1)$-dimensional Gross-Pitaevskii equation
\begin{align}
  \I \partial_t \hat\Phi=-\frac{1}{2 m} \nabla^2 \hat\Phi+V\,\hat\Phi+g_{2d} \hat\Phi|\hat\Phi|^2  \,,
\end{align}
where $\hat\Phi(t,\bm{x})$ is a complex scalar field, $m$ is the atomic mass of the BEC species, $g_{2d}$ is the $s$-wave scattering strength, and $V$ is a potential that we will take to be isotropic. Assuming a static BEC background $\Phi_0\in\mathbb{R}$ of infinite spatial extent, setting $V+g_{2d}\Phi_0^2=0$, and taking the long-wavelength limit, one finds that fluctuations about the BEC background obey the $(2+1)$ massless Klein-Gordon equation with propagation speed $c_s=\sqrt{g_{2d}\rho_0 /m}$, where $\rho_0=\Phi_0^2$ is the two-dimensional number density of the BEC background~\cite{PhysRevLett.85.4643,Unruh:2022gso}. The specific analogy we focus on here is given by
\begin{align}
   \hat \phi(t,\bm{x})\equiv \frac{1}{2\sqrt{m\rho_0}} \delta \hat\rho(t,\bm{x})\, ,
\end{align}
where $\hat\phi(t,\bm{x})$ is a massless Klein-Gordon field and $\delta \hat\rho(t,\bm{x})$ represents density fluctuations in the BEC \cite{Analog2}.

\begin{figure}
    \centering
    \includegraphics[width=\columnwidth]{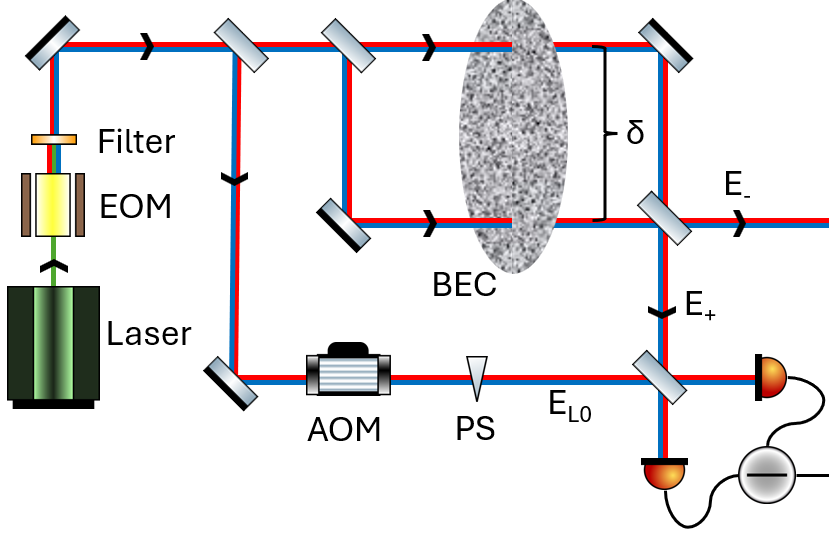}
    \caption{Experimental schematic for probing BEC density fluctuations from a superposition of detector locations separated by $\delta=|\bm{x}_1-\bm{x}_2|$. Prior to interaction, an electro-optic modulator (EOM) and a filter are used to prepare the modulated laser field \eqref{eqn:: plus minus decomp}. Post-interaction, the laser probes are recombined in a beamsplitter, and the ``sum'' output $E_+$ is heterodyned using a reference beam $E_{\mrm{LO}}$, yielding the desired response function \eqref{totalresponse2}. An acousto-optic modulator (AOM) and a phase-shifter (PS) are incorporated in the reference beam to facilitate the heterodyne process \cite{Analog2}.}
    \label{fig:exp}
\end{figure}

Using a highly-focused laser beam and a beamsplitter to produce a pair of spatially separated, superposed probes directed perpendicular to the BEC plane (as shown in Figure  \ref{fig:exp}), local BEC density fluctuations are transduced locally from the pair of interaction points into the laser phase. Suppressing spatial arguments for brevity, the quantum electric field for each of the two probes can be expressed as
\begin{align}\label{eqn:: Efield}
\hat E_j(t)~=~ \int \frac{\D\omega}{2\pi}\, E_0(\omega)\hat a_j[\omega]\E^{-\I\omega t}+\mrm{h.c.}
\end{align}
for $j\in \{1,2\}$, where $E_0(\omega)=\sqrt{ \omega /4\pi\varepsilon_0 A_\perp}$ (in natural units), $\varepsilon_0$ is the permittivity of free space, $A_\perp$ is the cross-sectional beam area, and the individual mode operators obey the commutation relation 
\begin{align}
\left[\hat a_j[\omega],\hat a_j^\dagger[\omega']\right]=2\pi\delta(\omega-\omega')\, .
\end{align}

Interaction with a laser field peaked at a single frequency produces an unbalanced force on the BEC, due to the Stark potential; to balance this force, a modulated laser field can be used, with two modulation peaks prepared on opposite sides of an atomic resonance \cite{Analog1}. For an atomic resonance at $\omega_0$ with upper $(+)$ and lower $(-)$ modulation bands peaked at $\omega_\pm=\omega_0\pm \Omega$ such that $\Omega\ll \omega_0$, we can represent the appropriate modulated laser field for each of the superposed probes with the mode operator decomposition
\begin{subequations}\label{eqn:: plus minus decomp}
\begin{align}
    \hat E_j(t)~=~&E_0(\omega_0)(\hat a_j(t)+\hat a_j^\dagger(t))\,,\\
    \hat a_j(t)~=~&\hat a_{j+}(t)+\hat a_{j-}(t)\,,\\
    \hat a_{j\pm}(t)~=~&(\alpha+\delta \hat a_{j\pm}(t))\E^{-\I\omega_\pm t}\,,
\end{align}    
\end{subequations}
where $\alpha\in\mathbb{R}$ is the coherent amplitude at each modulation band and we have evaluated the prefactor $E_0$ at $\omega_0$. Here $\delta \hat a_{j\pm}(t)$ are fluctuation operators that are slowly-varying with respect to the peak frequency $\omega_0$.

After interaction with the BEC, the upper and lower modulation band fluctuation operators are given by
\begin{equation}\label{delta_atilde}
\delta\hat a_{j\pm}(t)\rightarrow\delta \hat a_{j\pm}(t)\pm\frac{\I \mu \hat\phi(t,\bm{x}_j)}{\sqrt{2}} \pm \frac{\mu^2}{4}\delta \hat a_{jB}(t)\,,
\end{equation}
with the nonperturbative backaction expressed in the frequency domain as
\begin{align}
    &\delta \hat a_{jB}[\nu]=\int\D t\, \E^{\I\nu t} \delta \hat a_{jB}(t) \\
    &=\text{sgn}(\nu)\left( \delta \hat a_{j-}[\nu]+\delta \hat a_{j-}[-\nu]^\dagger-\delta \hat a_{j+}[\nu]-\delta \hat a_{j+}[-\nu]^\dagger\right)\nonumber.
\end{align}
Here $\mu$ is a dimensionless effective coupling parameter,
\begin{equation}
\mu=-|\alpha_R|\omega_0\sqrt{2\,m\,\rho_0}\alpha \, ,
\end{equation}
where $\alpha_R$ is the real part of the atomic polarizability \cite{Analog2}.

\section{Laser Detection Scheme}

% The laser-BEC interactions yield, at first order,
% \begin{equation}\label{eqa:firstorder}
%     \delta\tilde{a}_\pm(t)~=~\delta a_\pm(t)\mp\frac{1}{2}\I\varepsilon\alpha\phi(t,\bm{x}_a)\,,
% \end{equation}
% and
% \begin{equation}\label{eq:firstorder}
%     \delta\tilde{b}_\pm(t)~=~\delta b_\pm(t)\mp\frac{1}{2}\I\varepsilon\alpha\phi(t,\bm{x}_b)\,,
% \end{equation}
% for the two branches of the superposition. 

The signal we seek is the total response function 
\begin{align}\label{totalresponse}
    \mathcal{F}(\nu)=\mathcal{F}_{11}(\nu)+\mathcal{F}_{12}(\nu)+\mathcal{F}_{21}(\nu)+\mathcal{F}_{22}(\nu)\, ,
\end{align}
where, for $i,j\in \{1,2\}$, 
\begin{align}
    \mathcal{F}_{ij}(\nu)=\int\D t\, e^{-\I\nu t}\bra{0} \hat\phi(t,\bm{x}_i)\hat\phi(0,\bm{x}_j)\ket{0}
\end{align}
denote the diagonal ($i=j$) and off-diagonal ($i\neq j$) response functions. Reinstating the propagation speed $c_s$, it follows from \eqref{singlimit} and \eqref{offdiagrespo} that these response functions are given by
\begin{align}
    \mathcal{F}_{11}(\nu)=\mathcal{F}_{22}(\nu)=\frac{1}{2}\Theta(-\nu)
\end{align}
and 
\begin{align}
    \mathcal{F}_{12}(\nu)=\mathcal{F}_{21}(\nu)=\frac{1}{2}\Theta(-\nu)J_0(\nu|\bm{x}_1-\bm{x}_2|/c_s)\,, 
\end{align}
respectively. Hence, the total response function is
\begin{align}\label{totalresponse2}
    \mathcal{F}(\nu)=\Theta(-\nu)\left(1+J_0(\nu|\bm{x}_1-\bm{x}_2|/c_s)\right)\, .
\end{align}

The superposed signal-carrying laser fields \eqref{delta_atilde} are recombined in a beamsplitter, yielding the ``sum'' and ``difference'' outputs
\begin{align}\label{outputs}
    \hat E_\pm(t)=\frac{1}{\sqrt{2}}\left(\hat E_1(t)\pm \hat E_2(t)\right)\, .
\end{align}
Heterodyning the ``sum'' output $\hat E_+(t)$ yields the difference-photocurrent power spectral density (PSD) 
\begin{align}\label{diffPSDinf0}
S_{ii}[\Delta_{\mrm{LO}}-\nu]=     \frac{\mu^2}{2}\left(\mathcal{F}(\nu)+\mathcal{N}[\nu;\mu^2]\right)\, ,
\end{align}
where $\Delta_{\mrm{LO}}$ is the heterodyne shift frequency, the ``added noise'' $\mathcal{N}[\nu;\mu^2]$ is given by
\begin{align}\label{diffPSDadd0}
   \mathcal{N}[\nu;\mu^2]=\frac{1}{\mu^2}+\frac{3\mu^2}{8}+\sgn(\nu)
\end{align}
and $\mathcal{F}(\nu)$ corresponds to the total response function \eqref{totalresponse2} \cite{Analog2}. The first term on the right-hand-side of \eqref{diffPSDadd0} is noise from measurement imprecision (i.e. shot noise), the second arises from backaction-backaction correlations, and the third term comes from backaction-imprecision correlations. 

The standard quantum limit corresponds to the minimum of the added noise \eqref{diffPSDadd0}, which occurs for $\mu^2=\mu_{\text{SQL}}^2=2\sqrt{2/3}$, in which case the added noise is $\mathcal{N}[\nu;\mu_{\text{SQL}}^2]=\sqrt{3/2}+\text{sgn}(\nu)$. Using a particular class of squeezed initial states for the laser field, it is possible to beat the standard quantum limit for detection of the de-excitation spectrum ($\nu<0$) by nearly $15\%$ with only a modest amount of squeezing ($\sim 3$dB) \cite{Analog2}. Estimating the signal-to-noise ratio in this case, for frequencies $-c_s/|\bm{x}_1-\bm{x}_2|\lesssim \nu <0$, one finds
\begin{align}
    \text{SNR}=\frac{\mathcal{F}}{\mathcal{N}}=\frac{1+J_0(\nu|\bm{x}_1-\bm{x}_2|/c_s)}{\frac{1}{6}(\sqrt{10}-2)}\gtrsim 10\, .
\end{align}

\section{Discussion}

The above considerations pertain to an experimental proposal to extract a signal of the de-excitation spectrum for a UdW detector in a superposition of locations interacting with an effective relativistic field in $(2+1)$ dimensions. The extracted signal corresponds to a detector response function, which includes contributions from interference between two branches of a superposition of detector locations. 

The proposed experiment could be realized under the conditions found to be both ideal and feasible for observing an analogue of the circular motion Unruh effect \cite{Analog2}. The specific parameter regime corresponds to a BEC made from $^{133}$Cs atoms with a background density $\rho_0=10^3\mu\mrm{m}^{-2}$, and a laser of frequency $\omega_0/(2\pi)=10^{14}\mathrm{Hz}$ with a beam width $r_0=3\mu\mathrm{m}$. The laser power required to reach (and surpass) the standard quantum limit was also shown to ensure a nondestructive continuous measurement of the BEC density fluctuation field. The same conditions are ideal for an implementation of our proposal here. 

The response function is extracted by using a laser field as a quantum sensor \cite{RevModPhys.89.035002}. On the other hand, the response function characterizes the behaviour of a continuum of UdW detectors, each of which can itself be interpreted as a quantum sensor, as it is a qubit with a quantum control degree of freedom governing the qubit location. An analysis of the relevant quantum Fisher information for further experimental optimization is the subject of future research.

Included in the proposed response function signal \eqref{totalresponse2} are contributions from the responses along each individual trajectory, in addition to the off-diagonal responses. It may be appropriate, then, to consider an alternative signal, made by combining the ``sum'' heterodyne results with those obtained by heterodyning the ``difference'' output $\hat E_-(t)$. The latter is associated with the altered signal
\begin{align}\label{sigalt}
    \mathcal{F}_{\text{diff}}(\nu)=\Theta(-\nu)\left(1-J_0(\nu|\bm{x}_1-\bm{x}_2|/c_s)\right)\, .
\end{align}
The heterodyne difference-photocurrent for the ``difference'' output in this case would have the PSD
\begin{align}\label{diffPSDinf1}
S_{ii}^{\text{diff}}[\Delta_{\mrm{LO}}-\nu]=     \frac{\mu^2}{2}\left(\mathcal{F}_{\text{diff}}(\nu)+\mathcal{N}[\nu;\mu^2]\right)\, .
\end{align}
Subtracting the ``difference'' PSD \eqref{diffPSDinf1} from the ``sum'' output heterodyne PSD \eqref{diffPSDinf0} leads to a cancellation between common-trajectory (i.e. diagonal) contributions to the response function, while retaining the nonlocal (i.e. off-diagonal) contributions. The result isolates an interference signal: it is only nonvanishing when a superposition is present, and therefore could serve as a witness for such detector superpositions.   

The experiment discussed here is perhaps the simplest one that could be performed that would test foundational issues of physics associated with quantum reference frames \cite{Giacomini2019}. Once carried out, a range of future applications will become available, including testing
ideas about   causality
\cite{Henderson:2020zax}
and 
thermality \cite{Foo:2020xqn,Martin-Martinez:2010gnz},  simulating spacetime superposition \cite{Foo:2022dnz,Goel:2024vtr}, and 
implementing 
operational approaches to quantum gravity \cite{Foo:2020jmi,Foo:2021exb,Suryaatmadja:2023onb}.

% \vspace{14pt}

\section{Acknowledgements} CG wishes to thank Massimo Frigerio and Nicolas Treps for helpful discussions. CG acknowledges support provided by the European Research Council under the Consolidator Grant COQCOoN (Grant No. 820079).  This work was supported in part by the Natural Sciences and Engineering Research Council of Canada.

For the purpose of open access, the authors have applied a CC BY public copyright licence to any Author Accepted Manuscript version arising.  

\bibliography{refs.bib}

%\begin{appendix}

%\end{appendix}

\end{document}